\documentclass[12pt,eqsecnum]{article}
\usepackage{graphicx}
\usepackage[dvips]{color}
\usepackage{amssymb}
\usepackage{amsmath}
\usepackage{ulem}

\usepackage[dvipsnames]{xcolor}

\usepackage{soul}
\usepackage{ascmac}
\usepackage{cite}

\usepackage{amsfonts,amssymb,ascmac,theorem,url,ulem}

\makeatletter

\@addtoreset{equation}{section}
\makeatletter

\newcommand{\D}{{\rm d}}

{\theorembodyfont{\upshape}
}
{\theorembodyfont{\upshape}
}
{\theorembodyfont{\upshape}
}
{\theorembodyfont{\upshape}
}
{\theorembodyfont{\upshape}
}
{\theorembodyfont{\upshape}
}

\newcommand{\dalm}{\kern1pt\vbox{\hrule height 0.9pt\hbox{\vrule width
0.9pt\hskip 2.5pt\vbox{\vskip 5.5pt}\hskip 3pt\vrule width 0.3pt}\hrule height
0.3pt}\kern1pt}

\begin{document}

\begin{titlepage}
\vfill
\begin{flushright}
\today
\end{flushright}

\vfill
\begin{center}
\baselineskip=12pt
{\Large\bf
Extremal rotating BTZ black holes cannot be dressed \\
in (anti-)self-dual Maxwell field
}
\vskip 0.0cm
{\large {\sl }}
\vskip 0.mm
{\bf Hideki Maeda${}^{a}$ and Ji\v{r}\'{\i} Podolsk\'y$^{b}$} \\

\vskip 0.5cm
{
${}^a$Department of Electronics and Information Engineering, Hokkai-Gakuen University, Sapporo 062-8605, Japan.\\
${}^b$Institute of Theoretical Physics, Charles University, V~Hole\v{s}ovi\v{c}k\'ach 2, 18000 Prague 8, Czechia.\\
\texttt{h-maeda@hgu.jp, jiri.podolsky@mff.cuni.cz}
}
\vspace{0pt}
\end{center}
\vskip 0in
\par
\begin{center}
{\bf Abstract}
\end{center}
\begin{quote}
Under the (anti-)self-dual condition for orthonormal components of the Faraday tensor, the three-dimensional Einstein-Maxwell system with a negative cosmological constant $\Lambda$ admits a solution obtained by Kamata and Koikawa and later by Cataldo and Salgado in the most general form. Actually, Cl\'ement first obtained this solution and interpreted it as a regular particle-like solution without horizon. Nevertheless, it has been erroneously stated in some literature that this Cl\'ement-Cataldo-Salgado (CCS) solution, locally characterized by a single parameter, describes a black hole even in the charged case as it reduces to the extremal rotating Ba\~nados-Teitelboim-Zanelli (BTZ) solution in the vacuum limit and its curvature invariants are constant.
In this paper, we supplement Cl\'ement's interpretation by showing that there appears a parallelly propagated curvature singularity corresponding to an infinite affine parameter along spacelike geodesics at the location of the Killing horizon in the extremal rotating BTZ solution when the (anti-)self-dual Maxwell field is added. If the spatial coordinate $\theta$ is periodic, closed timelike curves exist near the singularity. It is also shown that the CCS solution is of the Cotton type~N (in contrast to charged rotating BTZ black holes which are of type~I away from the horizon), and the energy-momentum tensor of the Maxwell field is of the Hawking-Ellis type II. The metric solves the Einstein-$\Lambda$ equations also with a massless scalar field or a null dust fluid. We explicitly demonstrate that it belongs to the Kundt shear-free, non-twisting, and non-expanding class of geometries, whereas extremal rotating BTZ black holes have expanding principal null directions. It means that the CCS metric represents the specific null (that is ``radiative'') Maxwell field generated by a singular source, rather than an extremal rotating BTZ black hole dressed in an (anti-)self-dual Maxwell field.
\vfill
\vskip 2.mm
\end{quote}
\end{titlepage}




\tableofcontents

\newpage

\section{Introduction}

Three-dimensional gravity has been studied very intensively so far as a testing ground for quantum gravity due to its simplicity~\cite{Carlip:1998uc}. In particular, the Ba\~nados-Teitelboim-Zanelli (BTZ) vacuum black-hole solution~\cite{Banados:1992wn} in the presence of a negative cosmological constant $\Lambda$ is expected to provide clues to the description of black holes in quantum gravity. Since the number of independent components of the Riemann tensor and Ricci tensor are the same in three dimensions, the BTZ spacetime is locally identical to the maximally symmetric anti-de~Sitter (AdS) spacetime. Nevertheless, their global structures are different as the BTZ spacetime is obtained by the identifications in the AdS spacetime~\cite{Banados:1992gq}.

In the three-dimensional Einstein-Maxwell-$\Lambda$ system, the charged non-rotating BTZ solution was obtained in Ref.~\cite{Banados:1992wn} and independently in Ref.~\cite{Peldan:1992mp}.
Then, the charged rotating BTZ solution was obtained by linear coordinate transformations from the non-rotating solution by Cl\'ement~\cite{Clement:1995zt}. (See the introduction in~\cite{Maeda:2023oei} for the history of this solution.)
Therefore, the rotating and non-rotating charged BTZ solutions are locally identical, but again, they are globally different if one uses a periodic coordinate. The charged rotating BTZ solution for $\Lambda<0$, which corresponds to the Kerr-Newman-AdS solution in four dimensions, can be extended for a wider range of $\Lambda$ by coordinate transformations~\cite{Maeda:2023oei}, however, the extended solution for $\Lambda\ge 0$ does not describe a black hole due to Ida's theorem~\cite{Ida:2000jh}.

In fact, other exact solutions exist in the three-dimensional Einstein-Maxwell-$\Lambda$ system besides the charged rotating BTZ solution as summarized in Chapter~11 of the textbook~\cite{Garcia-Diaz:2017cpv}. Among them, there is a charged solution obtained by Kamata and Koikawa~\cite{Kamata:1995zu}, which {reduces} to the {\it extremal} rotating BTZ solution in the uncharged limit. Unlike the charged BTZ solution, orthonormal components of the Faraday tensor of the Kamata-Koikawa solution satisfy the so-called {\it self-dual} or {\it anti-self-dual} condition\footnote{This terminology differs from the usual one for the Maxwell field and will be explained in Sec.~\ref{sec:CS-sol}.} and the curvature invariants are constant. As a generalization of the Kamata-Koikawa solution, Cataldo and Salgado obtained the known most general stationary and axisymmetric solution under the (anti-)self-dual condition on the Maxwell field~\cite{Cataldo:1999fh}. The Cataldo-Salgado solution is characterized by four parameters and its curvature invariants are also constant. For some reason, only the Kamata-Koikawa solution is mentioned in Garc\'\i{}a-D\'\i{}az's textbook~\cite{Garcia-Diaz:2017cpv}, while the Cataldo-Salgado solution is not included.

In Ref.~\cite{Clement:1995zt}, Cl\'ement pointed out that the Kamata-Koikawa solution has previously been presented in Eq.~(29) in Ref.~\cite{Clement:1993kc}. In fact, as we will show in the present paper, the Cataldo-Salgado solution, including the Kamata-Koikawa solution as a special case, is also locally identical to the Cl\'ement solution. For this reason, we will refer to this solution as the Cl\'ement-Cataldo-Salgado solution (CCS) solution.
As described in some literature~\cite{Kamata:1995zu,Garcia-Diaz:2017cpv}, one might expect that the CCS solution describes a black hole since the curvature invariants are constant as the uncharged rotating BTZ solution.
In Ref.~\cite{Clement:1995zt}, Cl\'ement interpreted the CCS solution as a perfectly regular particle-like solution without horizon. However, in this work, we supplement Cl\'ement's interpretation by showing that there appears a parallelly propagated curvature singularity at the location of the Killing horizon in the extremal rotating BTZ solution if the (anti-)self-dual Maxwell field is added. We also clarify the specific Cotton type of the spacetime and the Hawking-Ellis type of the energy-momentum tensor.

In this paper, we first explain some mathematical tools and review the CCS solution and the charged rotating BTZ solution in Sec.~\ref{sec:review}.
Then, we study the CCS solution in detail in~Sec.~\ref{sec:main}. The summary of our main results and final remarks are given in the final section.
Appendix~\ref{app:Clement} shows that the Cataldo-Salgado solution and the Cl\'ement solution are locally identical. Appendix~\ref{app:CSKK} explains the parameters of the {Cataldo-Salgado} solution. Throughout this article, the signature of the Minkowski spacetime is $(-,+,+)$, and a prime denotes differentiation with respect to the argument. We adopt the units such that ${c=1}$, and the conventions of curvature tensors such as ${[\nabla _\rho ,\nabla_\sigma]V^\mu ={{\cal R}^\mu }_{\nu\rho\sigma}V^\nu}$ and ${{\cal R}_{\mu \nu }={{\cal R}^\rho }_{\mu \rho \nu }}$. We use ${\kappa:=8\pi G}$ instead of the gravitational constant $G$.

\section{Preliminaries}
\label{sec:review}

The action of the three-dimensional Einstein-Maxwell-$\Lambda$ system for the spacetime metric $g_{\mu\nu}$ and the $U(1)$ gauge field $A_\mu$ is given by
\begin{align}
\label{action}
S[g_{\mu\nu}, A_{\mu}]=&\int \D^3x\,\sqrt{-g}\,\biggl(\frac{1}{2\kappa}({\cal R}-2\Lambda)-\frac{1}{4}F_{\mu\nu}F^{\mu\nu}\biggl),
\end{align}
where ${\cal R}$ is the Ricci scalar and $F_{\mu\nu}:=\nabla_\mu A_\nu-\nabla_\nu A_\mu$ is the Faraday tensor satisfying an identity $\nabla_{[\rho} F_{\mu\nu]}\equiv 0$.
We have omitted the boundary term in the action for simplicity.
Variation of the action gives the field equations
\begin{align}
\label{EFE}
\begin{aligned}
G_{\mu\nu}+\Lambda g_{\mu\nu}&=\kappa \,T_{\mu\nu}, \\
\nabla_\nu F^{\mu\nu}&=0,
\end{aligned}
\end{align}
where $G_{\mu\nu}$ is the Einstein tensor and the energy-momentum tensor $T_{\mu\nu}$ for the Maxwell field is given by
\begin{align}
T_{\mu\nu}=&F_{\mu\rho}F_\nu^{~\rho}-\frac 14 g_{\mu\nu}F_{\rho\sigma}F^{\rho\sigma}. \label{Tab-Max}
\end{align}
Recently, the field equations in this system have been fully solved without imposing any spacetime symmetry~\cite{Podolsky:2021gsa}.

In this paper, we study stationary and axisymmetric solutions described in the coordinates $(t,r,\theta)$ by the following metric
\begin{align}
\label{metric}
&\D s^2=-\frac{r^2}{R^2}\,f\,\D t^2+\frac{\D r^2}{f}+R^2\biggl(\D\theta+\frac{h}{R^2}\,\D t\biggl)^2,
\end{align}
where ${f=f(r)}$, ${R=R(r)}$, and ${h=h(r)}$.
In this spacetime, ${\sqrt{-g}=r}$ holds, and a regular null hypersurface ${r=r_{\rm h}}$ defined by the condition $f(r_{\rm h})=0$ is a Killing horizon.
In the domain ${f>0}$, natural orthonormal basis one-forms $\{E_\mu^{(0)}, E_\mu^{(1)}, E_\mu^{(2)}\}$ in the coordinate system (\ref{metric}) are given by
\begin{align}
\label{basis-KK}
&E_\mu^{(0)}\,\D x^\mu=\frac{\sqrt{r^2f}}{R} \,\D t,\qquad E_\mu^{(1)}\,\D x^\mu=\frac{1}{\sqrt{f}} \,\D r,\qquad E_\mu^{(2)}\,\D x^\mu=R\,\biggl(\D\theta+\frac{h}{R^2}\,\D t\biggl),
\end{align}
of which contravariant components are
\begin{align}
\label{vecorbasis-KK}
\begin{aligned}
&E^{(0)\mu}\,\partial_\mu = \frac{1}{\sqrt{r^2 f}}\,\Big(\! -R \,\partial_t+ \frac{h}{R}\, \partial_\theta \Big),\qquad E^{(1)\mu}\,\partial_\mu = \sqrt{f}\,\partial_r,\qquad E^{(2)\mu}\,\partial_\mu = \frac{1}{R}\,\partial_\theta.
\end{aligned}
\end{align}
However, the coordinate system (\ref{metric}) does not cover the Killing horizons, as the metric and the natural basis \eqref{basis-KK} diverge for ${f=0}$. To properly investigate the geometrical and physical properties, it is necessary to investigate the metric more carefully~\cite{Maeda:2023oei}.

\subsection{Kinoshita-Gundlach-Bourg-Davey quasi-local mass and angular momentum}
For general axisymmetric spacetimes in three dimensions, Kinoshita-Gundlach-Bourg-Davey (KGBD) quasi-local mass $m$ and angular momentum $j$ are defined by~\cite{Kinoshita:2021qsv,Gundlach:2021six}
\begin{align}
m &:=\frac{\pi}{\kappa}\,\big(\!-\Lambda\,\psi_\mu \psi^\mu+K_\mu K^\mu\big),\label{def-QLM2}\\
j &:=\frac{1}{\kappa}\,\varepsilon^{\mu\rho\sigma}\psi_\mu\nabla_\rho\psi_\sigma\quad\Bigl(\,=-\frac{2\pi}{\kappa}\psi_\mu K^\mu\Bigl),\label{def-QLJ2}
\end{align}
where ${\psi^\mu=(\partial/\partial\theta)^\mu}$ is the Killing vector generating axisymmetry.
Here $K^\mu$ is the generalized Kodama vector defined in Ref.~\cite{Kinoshita:2021qsv,Gundlach:2021six} by
\begin{align}
&K^\mu:=-\frac12\,\varepsilon^{\mu\rho\sigma}\nabla_\rho \psi_\sigma,\label{def-K2}
\end{align}
where $\varepsilon_{\mu\rho\sigma}$ is totally anti-symmetric volume three-form.
The vector $K^\mu$ shares the same properties as the Kodama vector in ${n\,(\ge 3)}$ dimensions~\cite{Kodama:1979vn,Maeda:2007uu}. If $\psi^\mu$ is hypersurface-orthogonal, ${j=0}$ holds and then $m$ and $K^\mu$ reduce to the Misner-Sharp quasi-local mass~\cite{Misner:1964je,Maeda:2006pm} and the Kodama vector~\cite{Kodama:1979vn,Maeda:2007uu} in three dimensions ($n=3$), respectively.
In the spacetime (\ref{metric}), the generalized Kodama vector is given by
\begin{align}
K^\mu\partial_\mu=\frac{1}{2r}\left(2RR'\partial_t-h'\partial_\theta\right).\label{Kd-general}
\end{align}

For the rotating BTZ {\it vacuum} solution~\cite{Banados:1992wn}, $m$ and $j$ are constants and the metric is written as \cite{Maeda:2023oei}
\begin{align}
\label{rotatingBTZ-mj}
\begin{aligned}
&\D s^2=-f\,\D t\,^2+f^{-1}\D r^2+r^2\biggl(\D\theta-\frac{4Gj}{r^2}\,\D t\biggl)^2,\\
&f(r)=-\Lambda r^2-8Gm+\frac{(4Gj)^2}{r^2}.
\end{aligned}
\end{align}
Depending on the parameters $m$ and $j$, the spacetime admits {two Killing horizons}, at most. They are located at ${r=r_{\rm h}}$ determined by $f(r_{\rm h})=0$, namely, $r_{\rm h}=r_\pm$, where
\begin{align}
& r_\pm^2:= \frac{4Gm}{(-\Lambda)}\, \bigg( 1 \pm \sqrt{ 1 + \Lambda \frac{j^2}{m^2}}\,\, \bigg).
\end{align}
The {\it extremal} rotating BTZ vacuum solution is realized for
\begin{align}
|m|=\sqrt{-\Lambda}\,|j|. \label{extreme-mj}
\end{align}
In such a case the metric function becomes
\begin{align}\label{horizon}
&f(r)= (-\Lambda)\, \frac{(\,r^2-r^2_{\rm ex})^2}{r^2},
\end{align}
with (assuming ${j>0}$) one double-degenerate Killing horizon located at ${r=r_{\rm ex}}$, where
\begin{align}
r^2_{\rm ex} :=\frac{4Gj}{\sqrt{-\Lambda}} = \frac{4Gm}{(-\Lambda)}.
\end{align}

Recall that, in the rotating (${j\ne 0}$) BTZ vacuum solution, ${r=0}$ is a coordinate singularity and the spacetime can be analytically extended using the coordinate ${y:=r^2}$ into the region of ${y<0}$.
In the coordinates $(t,y,\theta)$, the metric (\ref{rotatingBTZ-mj}) is written as
\begin{align}
\label{rotating-BTZ-r2}
\begin{aligned}
\D s^2 &= -(-\Lambda y-M)\D t^2-8Gj\,\D t\,\D\theta+\frac{\D y^2}{4y{\bar f}}+y\,\D\theta^2 \\
&= -{\bar f}\,\D t^2+\frac{\D y^2}{4y{\bar f}}+y\biggl(\D\theta-\frac{4Gj}{y}\D t\biggl)^2,\\
{\bar f}(y)&:=-\Lambda y-M+\frac{(4Gj)^2}{y},
\end{aligned}
\end{align}
where ${M=8G m}$, which give the determinant of the metric ${\mbox{det}(g) =-1/4}$, and non-zero components of the inverse metric as
\begin{align}
g^{tt}=-\frac{1}{{\bar f}}, \qquad g^{t\theta}=-\frac{4Gj}{y{\bar f}},\qquad g^{yy}=4y{\bar f},\qquad g^{\theta\theta}=-\frac{\Lambda y+M}{y{\bar f}}.
\end{align}
Because the metric and its inverse are analytic at ${y(=r^2)=0}$ for $j\ne 0$, the spacetime in the region of ${y>0}$ can be analytically extended beyond ${y=0}$ into the region of ${y<0}$.
As ${g^{\mu\nu}(\nabla_\mu y)(\nabla_\nu y)|_{y=0}=g^{yy}|_{y=0}=4(4Gj)^2>0}$, ${y=0}$ is a regular timelike hypersurface.\footnote{In Ref.~\cite{Briceno:2024ddc}, the authors use the singular coordinates $(t,r,\theta)$ of \eqref{rotatingBTZ-mj} and claim that ${r=0}$ is singular based on the analysis of the holonomy operator on a closed path (i.e., Wilson loop) around ${r=0}$.}

\subsection{Cl\'ement-Cataldo-Salgado solution}
\label{sec:CS-sol}

Cataldo and Salgado obtained the known most general stationary and axisymmetric solution assuming a {\it self-dual} or {\it anti-self-dual} condition imposed on the orthonormal basis components of the electric and magnetic fields~\cite{Cataldo:1999fh}. It can be written in the form of the line element~(\ref{metric}) in the coordinates $(t,r,\theta)$ with the metric functions given by Eqs.~(40)--(42) in~\cite{Cataldo:1999fh}, namely
\begin{align}
\label{CSKK-functions}
\begin{aligned}
f(r)&=(-\Lambda)\, \frac{(r^2-D)^2}{r^2},\\
R(r)&= \sqrt{r^2+\frac{\kappa{C_0^2}}{2(-\Lambda)}\ln|r^2-D|}\,,\\
h(r)&= \epsilon\,
\Big( D \sqrt{-\Lambda} + \frac{\kappa{C_0^2}}{2\sqrt{-\Lambda}}\ln\big|r^2-D\big|\Big),
\end{aligned}
\end{align}
while the gauge field is
\begin{align}
\label{CSKK-A}
A_\mu\,\D x^\mu =
\frac{1}{2}\,C_0\,\ln|r^2-D|
\,\Big(\epsilon\,\D t + \frac{\D\theta}{\sqrt{-\Lambda}}\Big).
\end{align}
Using the useful identities
\begin{align}
\label{CSKK-functions-identities}
\begin{aligned}
\sqrt{r^2 f}&= \sqrt{-\Lambda}\, (r^2-D),\\
h+\epsilon\sqrt{r^2f} &= \epsilon\,\sqrt{-\Lambda}\, R^2,\\
h-\epsilon\sqrt{r^2f} &= \epsilon\,\sqrt{-\Lambda}\, R^2 -2\epsilon\sqrt{r^2f}
\end{aligned}
\end{align}
in the region of $r^2>D$, the metric can be rewritten in a simple form
\begin{align}
\label{metric-rttheta}
\D s^2 &= (-\Lambda)\big[R^2-2(r^2-D) \big]\,\D t^2 + 2\epsilon\sqrt{-\Lambda}\,\big[R^2-(r^2-D) \big]\, \D t\,\D \theta+R^2\D \theta^2 +\frac{\D r^2}{f} ,
\end{align}
and the inverse metric is given by
\begin{align}
\label{metric-rttheta-inverse}
\begin{aligned}
&g^{tt}=\frac{R^2}{\Lambda(r^2-D)^2},\qquad g^{t\theta}=\epsilon\frac{R^2-(r^2-D)}{\sqrt{-\Lambda}(r^2-D)^2},\\
&g^{\theta\theta}=-\frac{R^2-2(r^2-D)}{(r^2-D)^2},\qquad g^{rr}=f.
\end{aligned}
\end{align}
Non-zero components of the Maxwell field $F_{\mu\nu}$ and $F^{\mu\nu}$ are given by
\begin{align}
\label{F^ab}
\begin{aligned}
&F_{tr}=-\frac{\epsilon\,C_0\,r}{r^2-D},\qquad F_{r\theta}=\frac{C_0\,r}{\sqrt{-\Lambda}\,(r^2-D)},\\
&F^{tr}=\epsilon\,\frac{C_0}{r},\qquad\qquad\! F^{r\theta}=\sqrt{-\Lambda}\,\,\frac{C_0}{r},
\end{aligned}
\end{align}
and hence the main Maxwell electromagnetic invariant is vanishing,
\begin{align}
F_{\mu\nu}F^{\mu\nu} = 0.
\end{align}
As shown in Appendix~\ref{app:Clement}, this Cataldo-Salgado solution is locally identical to Cl\'ement's solution given by Eq.~(29) in Ref.~\cite{Clement:1993kc}, namely, one solution can be obtained by coordinate transformations from the other.
For this reason, we will refer to this solution as the Cl\'ement-Cataldo-Salgado (CCS) solution.

Apart from the (negative) cosmological constant~$\Lambda$, the CCS solution contains {\it two continuous real parameters}, $D$ and $C_0$, and also a {\it discrete parameter} ${\epsilon=\pm 1}$, introduced already in \cite{Kamata:1995zu, Cataldo:1999fh}.
In fact, the original form of the solution presented in~\cite{Cataldo:1999fh} includes two additional parameters $C$ and $E$, but we have set ${C=1}$ and ${E=0}$ by using the coordinate freedom and suitable redefinitions of the other parameters, as shown explicitly in Appendix~\ref{app:CSKK}. Although those additional parameters $C$ and $E$ possibly may have {\it global} meanings, in this paper we focus on the local properties of the solution.

Moreover, one can also set ${D=0}$ in the charged case ${C_0\ne 0}$ by coordinate transformations as shown in Appendix~\ref{app:CSKK}, so that $C_0$ is the {\it only} continuous parameter characterizing the CCS solution.
However, by setting ${D=0}$, we miss the limit from the CCS solution to the Kamata-Koikawa solution~\cite{Kamata:1995zu} for ${\kappa {C_0^2}=-2\Lambda D}$. For this reason, we will keep $D$ arbitrary in the following analysis.

Since the CCS solution {reduces} to the extremal rotating BTZ {\it vacuum} solution for ${C_0=0}$, the parameter $C_0$ is related to the {electric charge}~$Q_E$, while the parameter $D$ corresponds to the extreme horizon $r^2_{\rm ex}$ of the vacuum BTZ black hole in view of Eq.~\eqref{horizon}. In fact, ${C_0=Q_E}$ holds, as can be immediately seen from the asymptotic behavior ${r \to \infty}$ of the electric field component $F^{tr}$ given by Eq.~\eqref{F^ab} (provided $r$ is the radial distance from the charge). Therefore, the constant factor which occurs in \eqref{CSKK-functions} is actually ${\kappa{C_0^2} = 8\pi G\,Q_E^2}$.

On the other hand, the physical meaning of the discrete parameter ${\epsilon=\pm 1}$ is twofold.
First, since the metric (\ref{metric-rttheta}) shows that changing the sign of $\epsilon$ is equivalent to the transformation of time reversal ${t\leftrightarrow-t}$, the two possibilities ${\epsilon=\pm 1}$ represent the choice of the {time orientation}.
Second, it denotes two possible types of the electromagnetic field, namely,
\vspace{1mm}
\begin{itemize}
\item ${\epsilon=+1}$: {\it self-dual} Maxwell field,

\item ${\epsilon=-1}$: {\it anti-self-dual} Maxwell field,
\end{itemize}
which is exhibited by the frame components of the Faraday tensor ${F_{(a)(b)} :=F_{\mu\nu}\,E^\mu_{(a)}E^\nu_{(b)}}$.
With the orthonormal basis one-forms (\ref{basis-KK}) in the domain ${r^2>D}$, non-zero components of ${F_{(a)(b)}}$ are
\begin{align}
\label{EB}
\begin{aligned}
& F_{(0)(1)}=-F_{(1)(0)}=:{\cal E}, \\
& F_{(2)(1)}=-F_{(1)(2)}=:{\cal B},
\end{aligned}
\end{align}
where the electric component~${\cal E}$ and the magnetic component~${\cal B}$ of ${F_{(a)(b)}}$ are given by
\begin{align}
{\cal E} =\epsilon\,{\cal B} \,\,= -\epsilon\,\frac{C_0}{R}.\label{def-selfdual}
\end{align}
Hence, the {\it self-dual condition} ${\,{\cal E}={\cal B}\,}$ and the {\it anti-self-dual condition} ${\,{\cal E}=-{\cal B}\,}$ introduced in Ref.~\cite{Kamata:1995zu} are satisfied for ${\epsilon=1}$, and ${\epsilon=-1}$, respectively, and a self-dual solution is the time reversal of an anti-self-dual solution.

Using ${\psi^\mu\partial_\mu=\partial_\theta}$ and Eq.~(\ref{Kd-general}), which for \eqref{CSKK-functions} is
\begin{align}
K^\mu\partial_\mu = \partial_t + \frac{\kappa{C_0^2}}{2\sqrt{-\Lambda}\,|r^2-D|}
\Big(\frac{1}{\sqrt{-\Lambda}} \,\partial_t-\epsilon\,\partial_\theta\Big) ,
\end{align}
we compute the KGBD mass (\ref{def-QLM2}) and the KGBD angular momentum (\ref{def-QLJ2}) to obtain
\begin{align}\label{CSKK-m-and-j}
m=-\epsilon\,\sqrt{-\Lambda}\,j =\,\frac{1}{8G}\Bigl[\,2D\,(-\Lambda)-\kappa{C_0^2}
\big( 1 - \ln\big|r^2-D\big|\big)\Bigl].
\end{align}
While the CCS solution \eqref{CSKK-functions} reduces to the extremal rotating BTZ {\it vacuum} solution \eqref{rotatingBTZ-mj} for ${C_0=0}$, it satisfies the extremality condition (\ref{extreme-mj}) independent of the parameters $D$, $C_0$, and $\epsilon$.

\subsection{Charged rotating BTZ solution}

In this subsection, for comparison with the CCS solution we review the charged rotating BTZ solution.
In~Ref.\cite{Clement:1995zt}, Cl\'ement obtained the so-called charged rotating BTZ solution in the system (\ref{action}) with ${\Lambda<0}$. After suitable reparametrizations, the solution is written in the form of the line element~(\ref{metric}) with the metric functions given by
\begin{align}
\label{Clement-functions}
\begin{aligned}
&f(r)=-\Lambda r^2-M-\kappa Q^2\ln r,\\
&R(r)=\sqrt{r^2+\frac{\omega^2}{(-\Lambda)(1-\omega^2)}\big(M+\kappa Q^2\ln r\big)},\\
&h(r)=-\frac{\omega}{\sqrt{-\Lambda}\,(1-\omega^2)}\big(M+\kappa Q^2\ln r\big),
\end{aligned}
\end{align}
while the gauge field is
\begin{align}
A_\mu\D x^\mu=-\frac{Q}{\sqrt{1-\omega^2}}\ln r\left(\D t-\frac{\omega}{\sqrt{-\Lambda}}\,\D\theta\right).\label{Clement-A}
\end{align}
We refer to the solution given by Eqs.~(\ref{Clement-functions}) and (\ref{Clement-A}) as the charged rotating BTZ solution in the {\it Cl\'ement form}.
Non-zero components of the inverse metric are given by
\begin{align}
\label{metric-Clement-inverse}
\begin{aligned}
&g^{tt}=-\frac{\omega^2f+\Lambda r^2}{\Lambda r^2f(1-\omega^2)},\qquad g^{t\theta}=\frac{\omega(f+\Lambda r^2)}{\sqrt{-\Lambda}\,r^2f(1-\omega^2)},\\
&g^{\theta\theta}=\frac{f+\omega^2\Lambda r^2}{r^2f(1-\omega^2)},\qquad g^{rr}=f.
\end{aligned}
\end{align}
This solution is parameterized by three constants, namely $\omega\,(\ne \pm 1)$, $M$, and $Q$, and requires ${\Lambda<0}$ in order for the metric to be real. (A different parametrization for a wider range of $\Lambda$ has been presented in~\cite{Maeda:2023oei}.)
Since there is a curvature singularity at ${r=0}$ for ${Q\ne 0}$, the domain of $r$ is restricted to ${r\in(0,\infty)}$.

Non-zero components of the Maxwell field $F_{\mu\nu}$ and $F^{\mu\nu}$ are given by
\begin{align}
\label{F^ab-Clement}
\begin{aligned}
&F_{tr}= \frac{Q}{\sqrt{1-\omega^2}\,r},\qquad F_{r\theta}=\frac{\omega Q}{\sqrt{-\Lambda(1-\omega^2)}\,r},\\
&F^{tr}=-\frac{Q}{\sqrt{1-\omega^2}\,r},\qquad F^{r\theta}=\frac{\omega Q\sqrt{-\Lambda}}{\sqrt{1-\omega^2}\,r},
\end{aligned}
\end{align}
which give the following main Maxwell electromagnetic invariant as
\begin{align}
F_{\mu\nu}F^{\mu\nu}=-\frac{2Q^2}{r^2}.\label{Clement-F2}
\end{align}
We note that, although the gauge field $A_\mu$ and the Faraday tensor $F_{\mu\nu}$ become pure imaginary if a condition ${-1<\omega<1}$ is not satisfied, the energy-momentum tensor $T_{\mu\nu}$ remains real even in such a case.
With the orthonormal basis one-forms (\ref{basis-KK}), non-zero components of ${F_{(a)(b)}}$ are
\begin{align}
\begin{aligned}
{\cal E} &:= F_{(0)(1)}=-F_{(1)(0)} = \frac{Q}{\sqrt{1-\omega^2}\,R}, \\
{\cal B} &:= F_{(2)(1)}=-F_{(1)(2)} = -\omega\,\sqrt{\frac{f}{-\Lambda r^2}}\,\,{\cal E},
\end{aligned}
\end{align}
where $f(r)$ is given by Eq.~\eqref{Clement-functions}. As they do not satisfy the relation~\eqref{def-selfdual}, the charged rotating BTZ solution is not self-dual nor anti-self-dual.
We note that the solution is purely electric in the non-rotating limit $\omega\to  0$.
In contrast, the CCS solution reduces to the massless BTZ {\it vacuum} solution in the non-rotating limit given by $D\to 0$ and $C_0\to 0$.

The KGBD mass and the KGBD angular momentum are given by
\begin{align}
m &= \frac{1+\omega^2}{8G(1-\omega^2)}\biggl(\frac12r f'-f\biggl)+\frac{\pi Q^2}{2(1-\omega^2)}\biggl(1+\frac{\omega^2}{2\Lambda r}f'\biggl) \nonumber\\
&= \frac{1+\omega^2}{8G(1-\omega^2)}(M+8\pi G Q^2\ln r)-\frac{\pi \omega^2Q^2}{1-\omega^2}\biggl(1+\frac{2\pi G Q^2}{\Lambda r^2}\biggl),\label{m-Clement}\\
j &= \frac{\omega}{4G\sqrt{-\Lambda}(1-\omega^2)}\left(\frac12r f'-f\right) \nonumber\\
&= \frac{\omega}{4G\sqrt{-\Lambda}(1-\omega^2)}\left(M+8\pi G Q^2\ln r-4\pi G Q^2\right).\label{j-Clement}
\end{align}
We note that $m$ and $j$ are non-constant in general for ${Q\ne 0}$. We also observe that ${j=0}$ is possible for ${\omega=0}$ with ${Q\ne 0}$.

In Ref.~\cite{Martinez:1999qi}, the authors identified the global mass $M_{\rm global}$, charge $Q_{\rm global}$, and angular momentum $J_{\rm global}$ as conserved quantities by the Regge-Teitelboim Hamiltonian approach.
In the units such that $8\pi G=\kappa=1/2$ and $-\Lambda=1$, our parameters $M$ and $Q$ are identical to ${\tilde M}$ and ${\tilde Q}$ in Ref.~\cite{Martinez:1999qi}, respectively, and hence $M_{\rm global}$, $J_{\rm global}$, and $Q_{\rm global}$ are given by
\begin{align}
\label{charges}
\begin{aligned}
&M_{\rm global}=\frac{1}{1-\omega^2}\biggl((1+\omega^2)M-\frac12\omega^2Q^2\biggl),\\
&J_{\rm global}=\frac{2\omega}{1-\omega^2}\biggl(M-\frac14Q^2\biggl),\qquad Q_{\rm global}=\frac{Q}{\sqrt{1-\omega^2}}.
\end{aligned}
\end{align}
With $8\pi G=1/2$ and $-\Lambda=1$, the KGBD mass (\ref{m-Clement}) and the KGBD angular momentum (\ref{j-Clement}) become
\begin{align}
m &= 2\pi\biggl[\frac{1}{1-\omega^2}\biggl((1+\omega^2)M-\frac{1}{2}\omega^2Q^2\biggl)+\frac{1+\omega^2}{2(1-\omega^2)}Q^2\ln r+\frac{\omega^2Q^4}{16(1-\omega^2)r^2}\biggl],\\
j &= 2\pi\biggl[\frac{2\omega}{1-\omega^2}\left(M-\frac14 Q^2\right)+\frac{\omega}{1-\omega^2} Q^2\ln r\biggl].
\end{align}
Interestingly, up to the normalization factor ${2\pi}$, the {\it constant terms} in the right-hand sides are {\it exactly} $M_{\rm global}$ and $J_{\rm global}$, respectively.

Cl\'ement's charged rotating BTZ solution (\ref{Clement-functions}) admits an {\it extremal Killing horizon} if $M$ and $Q$ satisfy the relation
\begin{align}
M = 4\pi G Q^2\biggl[ 1-\ln \biggl(\frac{4\pi G Q^2}{-\Lambda}\biggl)\biggl].\label{extreme-Clement}
\end{align}
The location of the extremal horizon ${r=r_{\rm ex}}$ is determined by ${f(r_{\rm ex})=f'(r_{\rm ex})=0}$ such as
\begin{align}
r_{\rm ex}=\sqrt{\frac{4\pi G Q^2}{-\Lambda}}.\label{r-ex}
\end{align}
Evaluating Eqs.~(\ref{m-Clement}) and (\ref{j-Clement}) on $r=r_{\rm ex}$, we obtain
\begin{align}
m(r_{\rm ex})=&\frac{\pi Q^2}{2(1-\omega^2)},\qquad j(r_{\rm ex})=0.\label{mj-Clement-ex}
\end{align}
In particular, the KGBD quasi-local angular momentum $j$ vanishes on the extremal horizon.

Unfortunately, the charged rotating BTZ solution in the Cl\'ement form (\ref{Clement-functions}) {\it cannot treat the extremal case in vacuum}.
In the uncharged case ${Q=0}$, the metric (\ref{metric}) with Eq.~(\ref{Clement-functions}) becomes
\begin{align}
\label{chargedBTZ-Q=0-F}
\begin{aligned}
\D s^2 &= -F\,\D t^2+F^{-1}\D R^2+R^2\biggl(\D\theta-\frac{M\omega}{\sqrt{-\Lambda}\,(1-\omega^2)R^2}\,\D t\biggl)^2,\\
F(R) &:= \frac{r^2}{R^2}\,f=-\Lambda R^2-\frac{1+\omega^2}{1-\omega^2}M+\frac{M^2\omega^2}{(-\Lambda)\,(1-\omega^2)^2R^2} \\
&\ =\frac{[M+\Lambda(1-\omega^2)R^2][M\omega^2+\Lambda(1-\omega^2)R^2]}{(-\Lambda)(\omega^2-1)^2R^2}
\end{aligned}
\end{align}
in the coordinates ${(t,R,\theta)}$, which is locally maximally symmetric and identical to the metric (\ref{rotatingBTZ-mj}) with Eq.~(\ref{mj-unchargedClement}).

The solution admits two Killing horizons (at most) in the region ${y>0}$ (recall that ${y :=R^2}$), and their locations ${y=y_1}$ and ${y=y_2}$ are the roots of ${F(R)=0}$, namely
\begin{align}
y_1=\frac{M\omega^2}{(-\Lambda)(1-\omega^2)}, \qquad y_2=\frac{M}{(-\Lambda)(1-\omega^2)}.
\end{align}
As ${\omega^2\ne 1}$ is assumed in the Cl\'ement form, the extremal case ${y_1=y_2}$ cannot be treated.

Let us show this fact differently. In the uncharged case ${Q=0}$, the KGBD mass (\ref{m-Clement}) and angular momentum (\ref{j-Clement}) for Cl\'ement's charged rotating BTZ solution reduce to the following constants
\begin{align}
m =\frac{1+\omega^2}{8G(1-\omega^2)}M,\qquad
j =\frac{M\omega}{4G\sqrt{-\Lambda}\,(1-\omega^2)}.\label{mj-unchargedClement}
\end{align}
The above equations are solved for $M$ and $\omega$ to give
\begin{align}
M=\mp 8G\sqrt{m^2+\Lambda j^2},\qquad
\omega=\frac{m\pm\sqrt{m^2+\Lambda j^2}}{\sqrt{-\Lambda}\,j}. \label{Momega-mj}
\end{align}
Under the extremality condition (\ref{extreme-mj}), namely ${m=\pm \sqrt{-\Lambda}\,j}$, we obtain
\begin{align}
M=0,\qquad \omega=\pm 1,
\end{align}
where the latter is not allowed in the Cl\'ement form. In fact, ${m}$ and ${j}$ are then {\it undetermined} in Eq.~(\ref{mj-unchargedClement}). Hence, the extreme case in vacuum cannot be treated properly under the parametrization (\ref{chargedBTZ-Q=0-F}).

This is also the case under a new parametrization of the charged rotating BTZ solution for a wider range of $\Lambda$ introduced in~\cite{Maeda:2023oei}.
In Ref.~\cite{Maeda:2023oei}, the gauge field and the metric functions are written as
\begin{align}
A_\mu\D x^\mu &=-Q\ln r\,(\D t-a\,\D\theta), \label{new-A}\\
R(r) &=\ \sqrt{\zeta r^2+a^2(M+\kappa Q^2\ln r)},\label{R-def}\\
f(r) &=-\Lambda r^2-M-\kappa Q^2\ln r,\label{f-chargedBTZ-def}\\
h(r) &=-a\,(M+\kappa Q^2\ln r),\label{h-def}
\end{align}
which are parametrized by $M$, $a$, and $Q$.
While $M$ and $Q$ are the same as in the Cl\'ement form, the rotation parameter $a$ and the constant $\zeta$ are related to $\omega$ as
\begin{align}
a=\frac{\omega}{\sqrt{-\Lambda}\,(1-\omega^2)},\qquad \zeta=\frac{1}{1-\omega^2}. \label{zeta-omega}
\end{align}
Those two constants satisfy
\begin{align}\label{zetaequation}
\zeta^2-\zeta+a^2\Lambda=0,
\end{align}
which shows
\begin{align}
\zeta=\frac12\left(1\pm\sqrt{1-4a^2\Lambda}\right).\label{zeta}
\end{align}
In this new parametrization, there are two branches of solutions depending on the sign in Eq.~(\ref{zeta}).
In the {uncharged} case (${Q=0}$) for ${\zeta\ne 0}$, one can write both branches of solutions in the same form as
\begin{align}
\label{chargedBTZ:Q=0}
\begin{aligned}
&\D s^2=-F\,\D {\bar t}\,^2+F^{-1}\D R^2+R^2\biggl(\D\theta-\frac{\zeta aM}{R^2}\D {\bar t}\biggl)^2,\\
&F(R)=-\Lambda R^2-\zeta(2\zeta-1)\,M+\frac{\zeta^2a^2M^2}{R^2},
\end{aligned}
\end{align}
where ${{\bar t}:=t/\zeta}$.
The solution admits two Killing horizons (at most) in the region of ${y>0}$, and their locations ${y=y_\pm}$ are roots of ${F(R)=0}$, namely
\begin{align}
y_\pm=\frac{\zeta M(2\zeta-1\pm 1)}{2(-\Lambda)},
\end{align}
where we have used Eq.~(\ref{zetaequation}). Clearly, the extremal case ${y_+=y_-}$ is not possible for ${\zeta M\ne 0}$.

In the uncharged case ${Q=0}$, the KGBD mass (\ref{m-Clement}) and angular momentum (\ref{j-Clement}) under the new parametrization become
\begin{align}
&m=\frac{\zeta(2\zeta-1)}{8G}M,\qquad j=\frac{\zeta }{4G}aM,
\end{align}
which give
\begin{align}
&m-\varepsilon\sqrt{-\Lambda}\,j=\frac{\zeta M}{8G}\,\big(2\zeta-1-2\varepsilon a\sqrt{-\Lambda}\,\big).
\end{align}
As the bracket in the right-hand side cannot be zero due to Eq.~(\ref{zeta}), the extremality condition (\ref{extreme-mj}), namely ${m=\varepsilon \sqrt{-\Lambda}j}$, with ${\varepsilon=\pm1}$, in the uncharged case is not satisfied for $\zeta M\ne 0$.
Hence, the {\it extremal case in vacuum cannot be treated in the new parametrization as well}.

\begin{figure}[htbp]
\begin{center}
\includegraphics[width=1.0\linewidth]{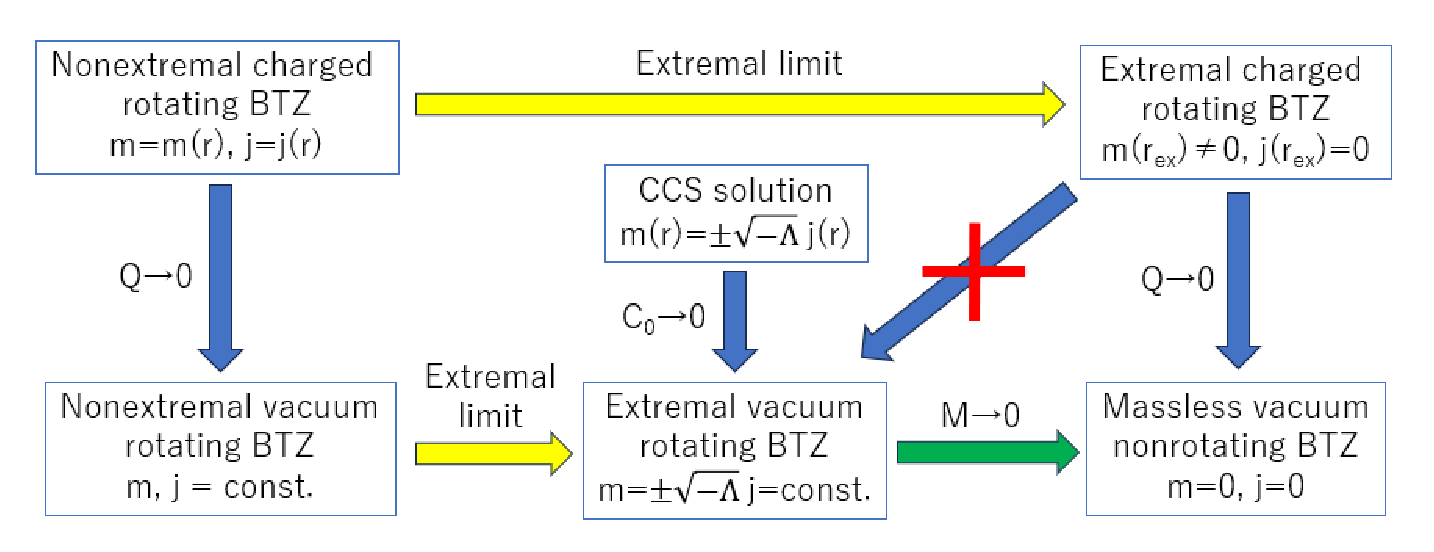}
\caption{\label{Fig-LimitingCases} Various limits from Cl\'ement's charged rotating BTZ solution, and from the CCS solution. }
\end{center}
\end{figure}

The results obtained in this subsection are summarized in Fig.~\ref{Fig-LimitingCases}.
We note that, unlike the extremal charged rotating BTZ solution in three dimensions, the extremal Kerr-Newman-AdS solution in four dimensions reduces to the extremal Kerr-AdS solution in the uncharged limit $Q\to 0$.

\section{Properties of the Cl\'ement-Cataldo-Salgado solution}
\label{sec:main}

In this section, we investigate geometrical and physical properties of the CCS solution in detail to give its physical interpretation.

\subsection{Geodesics and locally AdS infinity}

In the original coordinate system $(t,r,\theta)$ with the metric (\ref{metric-rttheta}), we consider the domain of $r$ given by ${r\in(0,\infty)}$, in which the metric component $g_{rr}$ or $g^{rr}$ diverges as ${r\to 0}$ or degenerates for any value of $D$. Actually, it is a coordinate singularity for $D\ne 0$ and can be removed by introducing a new radial coordinate $y:=r^2$.
In the new coordinates $(t,y,\theta)$, the metric (\ref{metric-rttheta}) and the gauge field (\ref{CSKK-A}) are written as
\begin{align}
\label{metric-y}
\begin{aligned}
\D s^2 & = (-\Lambda)\big[S-2(y-D) \big]\,\D t^2+ 2\epsilon\sqrt{-\Lambda}\,\big[S-(y-D) \big]\, \D t\,\D \theta \\
&\qquad\qquad +S\,\D \theta^2+ \frac{\D y^2}{4(-\Lambda)(y-D)^2} ,\\[1mm]
A_\mu\,\D x^\mu & =\frac{1}{2}\,C_0\,\ln|y-D|\,\Big(\epsilon\,\D t + \frac{\D\theta}{\sqrt{-\Lambda}}\Big),
\end{aligned}
\end{align}
where
\begin{align}
\label{def-S}
S(y):=y+\frac{\kappa{C_0^2}}{2(-\Lambda)}\ln|y-D|
\qquad\qquad \big[=R^2\big(r(y)\big)\,\big].
\end{align}
Non-zero components of the inverse metric are
\begin{align}
\label{metric-y-inverse}
\begin{aligned}
&g^{tt}=\frac{S}{\Lambda(y-D)^2},\qquad g^{t\theta}=\epsilon\frac{S-(y-D)}{\sqrt{-\Lambda}(y-D)^2},\\
&g^{\theta\theta}=-\frac{S-2(y-D)}{(y-D)^2},\qquad g^{yy}=4(-\Lambda)(y-D)^2.
\end{aligned}
\end{align}
Because for ${C_0 = 0}$ the metric and its inverse are both analytic at ${y=0}$ at ${D\ne 0}$, the spacetime defined in the domain ${y\ge 0}$ can be analytically extended beyond ${y=0}$ into the domain ${y<0}$ in that case.
In contrast, as shown in Sec.~\ref{Sec:singular} below, in the {\it non-vacuum case} for ${C_0\ne 0}$, ${y=D}$ is not a coordinate singularity but a {\it p.p. scalar curvature singularity}. Consequently, the domain of $y$ in the coordinate system (\ref{metric-y}) is ${y\in(D,\infty)}$.

For later use, we derive geodesic equations in the coordinates $(t,y,\theta)$.
Consider an affinely parametrized geodesic $\gamma$ represented as ${x^\mu=x^\mu(\lambda)}$ with its tangent vector ${v^\mu}$ (given by ${\D x^\mu/\D\lambda)}$, where $\lambda$ is an affine parameter along $\gamma$ in the coordinate system (\ref{metric-y}).
As the spacetime admits two Killing vectors, namely ${\xi^\mu=(\partial/\partial t)^\mu}$ and ${\Theta^\mu=(\partial/\partial \theta)^\mu}$, there are conserved quantities ${E:=-\xi_\mu v^\mu}$ and ${L:=\Theta_\mu v^\mu}$ along $\gamma$.
Using them and the normalization ${\varepsilon=v_\mu v^\mu}$, where ${\varepsilon=-1}$, $0$, and $1$, corresponds to timelike, null, and spacelike $\gamma$, respectively, we can write down the geodesic equations as
\begin{align}
\label{geodesic-eq-y}
\begin{aligned}
&{\dot t}=\frac{\epsilon \sqrt{-\Lambda}L(S-y+D)+ES}{(-\Lambda)(y-D)^2},\\
&{\dot y}^2=4\varepsilon(-\Lambda)(y-D)^2+4 (E+ \epsilon \sqrt{-\Lambda}L)
\big[(E+ \epsilon \sqrt{-\Lambda}L)S - 2\epsilon \sqrt{-\Lambda}L(y-D)\big],\\
&{\dot \theta}=\frac{-\epsilon \sqrt{-\Lambda}L\big[S-2(y-D)\big]-E(S-y+D)}{\epsilon \sqrt{-\Lambda}(y-D)^2},
\end{aligned}
\end{align}
where a dot denotes differentiation with respect to $\lambda$. (Notice also the difference between $\varepsilon$ and $\epsilon$.)

Since $\lim_{y\to \infty}R^{\mu\nu}_{~~\rho\sigma}=\Lambda(\delta^\mu_{~\rho}\delta^\nu_{~\sigma}-\delta^\mu_{~\sigma}\delta^\nu_{~\rho})$ is satisfied, the CCS spacetime is asymptotically (at least) locally AdS near the coordinate infinity $y\to \infty$.
In fact, null geodesics $(\varepsilon=0)$ with $E^2> -\Lambda L^2$ or $E=-\epsilon \sqrt{-\Lambda}L$ and timelike geodesics ($\varepsilon=-1$) cannot reach $y\to \infty$ because the right-hand side of the radial geodesic equation~(\ref{geodesic-eq-y}) becomes negative in the limit $y\to \infty$.
Along other geodesics, we obtain
\begin{align}
\label{geodesic-eq-y-infty1}
&{\dot y}^2\simeq \ \left\{
\begin{array}{ll}
4(E^2+\Lambda L^2)y & (\varepsilon=0, E^2> -\Lambda L^2)\\
8\kappa{C_0^2}E^2\ln|y-D|/(-\Lambda) & (\varepsilon=0, E=\epsilon \sqrt{-\Lambda}L\ne 0)\\
4 (-\Lambda)y^2 & (\varepsilon= 1)
\end{array}
\right.
\end{align}
near $y\to \infty$. In the first and the third cases, Eq.~(\ref{geodesic-eq-y-infty1}) is integrated to give
\begin{align}
y(\lambda)\simeq \ \ \left\{
\begin{array}{ll}
(E^2+\Lambda L^2)(\lambda-\lambda_0)^2 & (\varepsilon=0, E^2>-\Lambda L^2)\\
e^{2\sqrt{-\Lambda}\,(\lambda-\lambda_0)} & (\varepsilon=1)
\end{array}
\right.,
\end{align}
where $\lambda_0$ is a constant. Hence, $y\to \infty$ corresponds to an {\it infinite} affine parameter $\lambda\to \infty$. It is also true in the second case in Eq.~(\ref{geodesic-eq-y-infty1}) for $\varepsilon=0$ with $E=\epsilon \sqrt{-\Lambda}L\ne 0$ because the left-hand side blows up in the following inequality
\begin{align}
\lim_{y\to D}\int^y\frac{\D y}{y}<\lim_{y\to D}\int^y\frac{\D y}{\sqrt{\ln y}}.
\end{align}
Therefore, the asymptotically locally AdS region $y\to \infty$ corresponds to both {\it spacelike and null infinities}.

Lastly, we can also show that the infinity ${y\to \infty}$ is {\it causally timelike} by the conformal compactification of spacetime.
The line element of the CCS spacetime $({\cal M}_3,g_{\mu\nu})$ described by the metric $g_{\mu\nu}$ given by Eq.~(\ref{metric-y}) can be written as $\D s^2=\Omega^{-2}\D {\bar s}^2$ with a conformal factor $\Omega=S^{-1/2}$ that satisfies $\lim_{y\to\infty}\Omega= 0$.
Here $\D {\bar s}^2$ is the line element of the conformally compactified spacetime $({\cal {\bar M}}_3,{\bar g}_{\mu\nu})$ given by
\begin{align}
\begin{aligned}
&\D {\bar s}^2 =(\epsilon \D \theta+\sqrt{-\Lambda}\,\D t)\biggl[ \epsilon \D \theta+\sqrt{-\Lambda}\biggl(1-\frac{2\big(y(y_*)-D)}{S(y(y_*)\big)}\biggl) \D t\biggl] + \D y_*^2,\\
&y_*:=\int^y\frac{\D y}{2\sqrt{-\Lambda}(y-D)\sqrt{S}}.
\end{aligned}
\end{align}
In the asymptotically locally AdS region ${y\to \infty}$, $({\cal {\bar M}}_3,{\bar g}_{\mu\nu})$ which shares the same light-cone structure with $({\cal M}_3,g_{\mu\nu})$ is asymptotically flat,
\begin{align}
\D {\bar s}^2|_{y\to \infty}=-(-\Lambda)\D t^2+\D \theta^2 + \D y_*^2.
\end{align}
Such a boundary ${y\to \infty}$ is causally timelike because it corresponds to a finite value of $y_*$, as shown by
\begin{align}
\lim_{y\to \infty}y_*\simeq \lim_{y\to \infty}\int^y\frac{\D y}{2\sqrt{-\Lambda}\,y^{3/2}}
= C_\infty - \frac{1}{\sqrt{-\Lambda}}\lim_{y\to \infty}y^{-1/2}=C_\infty,
\end{align}
where $C_\infty$ is an integration constant.

\subsection{Curvature singularity at ${r^2(\equiv y) = D}$}
\label{Sec:singular}

Here we investigate the properties of ${y=D}$. We first show that ${y=D}$ is {\it causally null} by the conformal compactification of spacetime.
Since $S$ is negative near ${y=D}$, we write the line element of the CCS solution (\ref{metric-y}) as ${\D s^2=\Omega^{-2}\D {\bar s}^2}$, but now with a conformal factor ${\Omega=(-S)^{-1/2}}$ that also satisfies ${\lim_{y\to D}\Omega = 0}$.
The line element $\D {\bar s}^2$ of such conformally compactified spacetime is
\begin{align}
\begin{aligned}
&\D {\bar s}^2 =-(\epsilon \D \theta+\sqrt{-\Lambda}\D t)\biggl[\epsilon \D \theta+\sqrt{-\Lambda}\biggl(1-\frac{2\big(y(y_*)-D\big)}{S\big(y(y_*)\big)}\biggl) \D t\biggl] + \D y_*^2,\\
&y_*:=\int^y\frac{\D y}{2\sqrt{-\Lambda}(y-D)\sqrt{-S}}.
\end{aligned}
\end{align}
Near $y=D$, it reduces to
\begin{align}\label{auxi1}
\D {\bar s}^2|_{y\to D}\simeq -(\epsilon \D \theta+\sqrt{-\Lambda}\,\D t)^2 + \D y_*^2,
\end{align}
which is two-dimensional.
The boundary $y=D$ is causally null because it corresponds to $|y_*|\to \infty$, shown by
\begin{align}\label{auxi2}
\lim_{y\to D}|y_*|\simeq \lim_{y\to D}\biggl|\int^y\frac{\D y}{\sqrt{2\kappa{C_0^2}}(y-D)\sqrt{-\ln|y-D|}}\biggl|
= \lim_{y\to D}\sqrt{\frac{-2\ln|y-D|}{\kappa{C_0^2}}}\to \infty.
\end{align}

In Ref.~\cite{Clement:1995zt}, Cl\'ement showed that only a particular class of {\it spacelike} geodesics can reach ${y=D}$, and it corresponds to an {\it infinite affine distance}.
Because the right-hand side of Eq.~(\ref{geodesic-eq-y}) becomes negative as ${y\to D}$ unless ${E= -\epsilon \sqrt{-\Lambda}\,L}$ due to ${\lim_{y\to D}S\to -\infty}$, geodesics with ${E\ne -\epsilon \sqrt{-\Lambda}L}$ cannot reach ${y=D}$.
With ${E= -\epsilon \sqrt{-\Lambda}L}$, the geodesic equations (\ref{geodesic-eq-y}) for ingoing $\gamma$ reduce to
\begin{align}\label{dot-tthetay}
&{\dot t}=\frac{E}{(-\Lambda)(y-D)},\qquad
{\dot \theta}=-\frac{E}{\epsilon \sqrt{-\Lambda}(y-D)},\qquad
{\dot y}=-2\sqrt{\varepsilon(-\Lambda)}(y-D),
\end{align}
which show that only spacelike geodesics (${\varepsilon=1}$) with ${E= -\epsilon \sqrt{-\Lambda}\,L}$ can arrive at ${y=D}$.
Along it, ${R_{\mu\nu}v^\mu v^\nu=2 \Lambda}$ is satisfied, and the radial geodesic equation is integrated to give
\begin{align}
y(\lambda)=\ D+e^{-2\sqrt{-\Lambda}\,(\lambda-\lambda_0)},
\end{align}
where $\lambda_0$ is a constant. The above expression shows that ${y=D}$ corresponds to an infinite affine parameter ${\lambda\to \infty}$ along such a spacelike $\gamma$. For this reason, Cl\'ement concluded that the CCS solution is perfectly regular~\cite{Clement:1995zt}. However, we will show that ${y=D}$ is not a regular spacelike infinity but a {\it parallelly propagated curvature singularity}.

It is known that curvature singularities are classified into two main categories~\cite{Hawking:1973uf}. Although a scalar polynomial (s.p.) curvature singularity is usually examined, it may miss a parallelly propagated (p.p.) curvature singularity.
A s.p. curvature singularity is defined by the blowing up of a scalar, formed as a polynomial in the curvature tensor, such as the Ricci scalar ${\cal R}$ and the Kretschmann scalar ${\cal R}_{\mu\nu\rho\sigma}{\cal R}^{\mu\nu\rho\sigma}$.
On the other hand, a p.p. curvature singularity is defined by the blowing up of a component of the Riemann tensor in a parallelly propagated (pseudo-)orthonormal frame ${{\cal R}_{(a)(b)(c)(d)}:={\cal R}_{\mu\nu\rho\sigma}\,E^\mu_{(a)}E^\nu_{(b)}E^\rho_{(c)}E^\sigma_{(d)}}$ with basis vectors ${E^\mu_{(a)}}$ along a curve.
A s.p. curvature singularity is a p.p. curvature singularity but the latter is not always the former. (See Sec.~3 in~\cite{Ashley:2003tr}.)
In fact, the following curvature invariants of the CCS solution (\ref{metric-y}) are constant:
\begin{align}\label{someinvariants}
\begin{aligned}
&{\cal R}=6\Lambda,\qquad {\cal R}_{\mu\nu}{\cal R}^{\mu\nu}={\cal R}_{\mu\nu\rho\sigma}{\cal R}^{\mu\nu\rho\sigma}=12\Lambda^2,\\
&{\cal R}_\mu^{~\nu}{\cal R}_\nu^{~\rho}{\cal R}_\rho^{~\mu}=24\Lambda^3,\qquad (\nabla_\rho {\cal R}_{\mu\nu})(\nabla^\rho {\cal R}^{\mu\nu})=0.
\end{aligned}
\end{align}
Nevertheless, as shown below, $y=D$ in the CCS spacetime (\ref{metric-y}) is a p.p. curvature singularity.

We consider a spacelike geodesic given by \eqref{dot-tthetay} with ${E=0=L}$, of which the tangent vector is
\begin{align}
&v^\mu\partial_\mu=-2\sqrt{-\Lambda}(y-D)\,\partial_y.
\end{align}
We also introduce parallelly propagated basis vectors $\{E^\mu_{(0)}, E^\mu_{(1)}, E^\mu_{(2)}\}$ with ${E^\mu_{(2)}=v^\mu}$ along the geodesic as
\begin{align}
\begin{aligned}
&E_{(0)}^\mu\partial_\mu=\frac{1}{2\sqrt{-\Lambda}\,(y-D)^{3/2}}\left[-(S+y-D)\,\partial_t+\epsilon\sqrt{-\Lambda}\,(S-y+D)\,\partial_\theta\right],\\
&E_{(1)}^\mu\partial_\mu=\frac{1}{2\sqrt{-\Lambda}\,(y-D)^{3/2}}\left[-(S-y+D)\,\partial_t+\epsilon\sqrt{-\Lambda}\,[S-3(y-D)]\,\partial_\theta\right],\\
&E_{(2)}^\mu\partial_\mu=-2\sqrt{-\Lambda}\,(y-D)\,\partial_y,
\end{aligned}
\end{align}
which satisfy ${E^{(a)\mu} E_{(b)\mu}=\mbox{diag}(-1,1,1)}$ and ${E^\nu_{(2)}\nabla_\nu E^\mu_{(a)}=0}$ for ${a=0,1,2}$. Then, the following orthonormal components of the Riemann tensor diverge as ${y\to D}$ along $\eta$ for ${C_0 \ne 0}$:
\begin{align}
R_{(0)(2)(0)(2)}=-\Lambda+\frac{\kappa C_0^2}{y-D},\qquad
R_{(1)(2)(1)(2)}=\Lambda+\frac{\kappa C_0^2}{y-D},\qquad
R_{(0)(2)(1)(2)}=\frac{\kappa C_0^2}{y-D}.
\end{align}
Therefore, ${y=D}$ is not a spacelike infinity but a p.p. scalar curvature singularity corresponding to an infinite affine parameter $\lambda$.

To summarize, we have shown the following properties of ${y=D}$:
\begin{enumerate}
\item It is causally null.
\item Among all geodesics, only spacelike geodesics with ${E= -\epsilon \sqrt{-\Lambda}\,L}$ can reach ${y=D}$, corresponding to an infinite affine parameter.
\item Some components of ${\cal R}_{(a)(b)(c)(d)}$ blow up as ${y\to D}$ along a spacelike geodesic with ${E=0=L}$.
\end{enumerate}
By the property~2, ${y=D}$ is not a null infinity nor a timelike infinity.
By the property~3, ${y=D}$ is a spacelike infinity, and also a p.p. curvature singularity.
However, such a singularity may be harmless because (i) it is not naked for any observer in the region of ${y>D}$, and (ii) no free-falling causal observer arrives there.

In spite that no causal geodesic reaches ${y=D}$, causal {\it curves} could.\footnote{For example, in the five-dimensional vacuum spacetime obtained in Ref.~\cite{Lu:2008zs}, there exist timelike curves, corresponding to accelerated timelike observers, that arrive at a wormhole throat in spite that no causal geodesic arrives there.}
However, any causal curve ${t=t(y)}$ with constant $\theta$ does not reach ${y=D}$ with a finite value of $t$.
For such curves, from (\ref{metric-y}) we obtain
\begin{align}
\biggl(\frac{\D t}{\D y}\biggl)^2\ge \frac{1}{4(-\Lambda)^2(y-D)^2\,[-S+2(y-D)]} \label{dt/dy}
\end{align}
with equality holding for null curves. For a null curve, we get
\begin{align}
\lim_{y\to D}t\simeq \pm \lim_{y\to D}\sqrt{\frac{-2\ln|y-D|}{(-\Lambda)\kappa{C_0^2}}}\to \pm\infty.
\end{align}
Since the right-hand side of Eq.~(\ref{dt/dy}) blows up as ${y\to D}$, any causal curve with constant~$\theta$ does not reach ${y=D}$ with a finite value of~$t$.

Similarly, it can be shown that any causal curve ${\theta=\theta(y)}$ with constant $t$ does not reach ${y=D}$ with a finite value of $\theta$.
Indeed for such curves, we obtain
\begin{align}
\biggl(\frac{\D \theta}{\D y}\biggl)^2\ge \frac{1}{4(-\Lambda)(y-D)^2(-S)} \label{dtheta/dy}
\end{align}
with equality holding for null curves.
For a null curve, we obtain
\begin{align}
\lim_{y\to D}\theta\simeq \pm \lim_{y\to D}\sqrt{\frac{-2\ln|y-D|}{\kappa{C_0^2}}}\to \pm\infty.
\end{align}
Since the right-hand side of Eq.~(\ref{dtheta/dy}) blows up as ${y\to D}$, any causal curve with constant~$t$ does not reach ${y=D}$ with a finite value of $\theta$. Generally, there is no null curve ${y=y(t,\theta)}$ that arrives at ${y=D}$ with finite values of~$t$ and~$\theta$ because we obtain
\begin{align}
\lim_{y\to D}| \epsilon\,\theta + \sqrt{-\Lambda}\,t |\simeq \lim_{y\to D}\int^y\frac{\D y}{2\sqrt{-\Lambda}\,(y-D)\sqrt{-S}}
= \lim_{y\to D}\sqrt{\frac{-2\ln|y-D|}{\kappa{C_0^2}}}\to \infty
\end{align}
along such curves, cf. \eqref{auxi1}, \eqref{auxi2}.
We note that a curve satisfying $\epsilon\,\theta + \sqrt{-\Lambda}\,t=0$ is spacelike because $\D s^2=\D y^2/[4(-\Lambda)(y-D)^2]>0$ is satisfied along it.

The above facts (do not prove but strongly) suggest that there is no causal curve reaching ${y=D}$, and therefore the CCS spacetime cannot be extended along such curves. We thus conclude that the CCS solution with the non-vanishing (anti-)self-dual Maxwell field does {\it not} admit a (regular) horizon of a black hole (at ${y\equiv r^2=D}$), as opposed to the claim in \cite{Garcia-Diaz:2017cpv} that the solution describes an extremal black hole with mass, angular momentum, and electric charge.

\subsection{Causality violations}

Here we study causality in the CCS spacetime under the assumption that $\theta$ is a periodic coordinate.
First, we consider the standard identification ${(t,y,\theta)=(t,y,\theta+2\pi)}$ in the coordinate system (\ref{metric-y}).
Then, as ${g_{\theta\theta}<0}$ holds near the curvature singularity ${y=D}$ due to ${\lim_{y\to D}S\to -\infty}$, the Killing vector ${\Theta^\mu=(\partial/\partial \theta)^\mu}$ becomes timelike and therefore there are closed timelike curves near ${y=D}$.

In fact, under a different way of identification, the CCS spacetime admits closed null geodesics {\it everywhere}.
To show it, we introduce a null coordinate defined by
\begin{align}
\label{defining-u}
u := \sqrt{-\Lambda}\,t + \epsilon\,\theta
\end{align}
and write the CCS solution (\ref{metric-y}) in the new coordinates $(u, y,\theta)$ as
\begin{align}
\label{metric-yu}
\begin{aligned}
&\D s^2 = -\big[2(y-D)-S\big]\,\D u^2+2 \epsilon\,(y-D)\, \D u\,\D \theta + \frac{\D y^2}{4(-\Lambda)(y-D)^2},\\
&A_\mu\,\D x^\mu =\frac{\epsilon\,C_0}{2\sqrt{-\Lambda}} \ln|y-D|\,\D u.
\end{aligned}
\end{align}
Since $g^{\mu\nu}(\nabla_\mu u)(\nabla_\nu u)=g^{uu}=0$ holds, ${u=u_0}$ is a null hypersurface.
In fact, the constant $u_0$ labels privileged null hypersurfaces in the Kundt family, as shown in Eqs.~(\ref{basis-CS-y-triad}) and (\ref{prinipal-triad5}) below.
Also, as ${g_{\theta\theta}=0}$ holds, the Killing vector ${\Phi^\mu=(\partial/\partial \theta)^\mu\ (\ne\Theta^\mu)}$ is null everywhere.

As the new coordinate system (\ref{metric-yu}) admits the Killing vectors ${\zeta^\mu=(\partial/\partial u)^\mu}$ and ${\Phi^\mu=(\partial/\partial \theta)^\mu}$, ${{\bar E}:=-g_{\mu\nu}\zeta^\mu v^\nu}$, ${{\bar L}:=g_{\mu\nu}\Phi^\mu v^\nu}$, and ${\varepsilon=g_{\mu\nu}v^\mu v^\nu}$ are conserved along a geo\-desic~$\gamma$ with its tangent vector $v^\mu$, where ${\varepsilon=-1,0,1}$ corresponds to timelike, null, and spacelike $\gamma$.
Hence, geodesic equations for $\gamma$ are given by
\begin{align}
\begin{aligned}
&{\dot u}=\frac{\epsilon {\bar L}}{y-D},\\
&{\dot y}^2=4\varepsilon(-\Lambda)(y-D)^2 - 4(-\Lambda){\bar L} \big[2({\bar L}-\epsilon {\bar E})(y-D)-{\bar L}S\big],\\
& {\dot \theta}=\frac{-\epsilon {\bar E}(y-D)+{\bar L}\big[2(y-D)-S\big]}{(y-D)^2}.
\end{aligned}
\end{align}
For null geodesics (${\varepsilon=0}$) with ${{\bar L}=0}$, the above geodesic equations are easily integrated to give
\begin{align}
\label{geodesic-eq-special}
u=u_0,\qquad y=y_0,\qquad {\theta}=-\frac{\epsilon {\bar E}}{y_0-D}\,\lambda+\theta_0,
\end{align}
where $u_0$, $y_0$, and $\theta_0$ are constants.
Hence, if we identify $(u,y,\theta)=(u,y,\theta+2\pi)$ in the coordinate system (\ref{metric-yu}), the CCS spacetime admits closed null geodesics {\it everywhere}, which are described by Eq.~(\ref{geodesic-eq-special}) with ${\bar E}\ne 0$.

Such causal pathology is circumvented by the re-interpretation of the CCS solution in terms of the related (Kundt-type) coordinate ${r_{\rm\!_K}}$, which is introduced below in \eqref{r-Kundt} and is non-cyclic. Actually, it is an affine parameter along the privileged quadruple degenerate Cotton-aligned null direction ${k}^\mu$, as shown in Eq.~(\ref{prinipal-triad5}) below.

\subsection{Cotton and Maxwell algebraic types}

Next, we determine the Cotton type of the CCS spacetime (\ref{metric-y}), using the method recently developed in~\cite{Podolsky:2023qiu,Papajcik:2023zen}. It uses five real Cotton scalars~$\Psi_{\rm A}~(A=0,1,\cdots,4)$ defined by
\begin{align}
\label{Cottonscalars}
\begin{aligned}
&\Psi_0:=C_{\mu\nu\rho}\,k^\mu m^\nu k^\rho, \\
&\Psi_1:=C_{\mu\nu\rho}\,k^\mu l^\nu k^\rho,\\
&\Psi_2:=C_{\mu\nu\rho}\,k^\mu m^\nu l^\rho,\\
&\Psi_3:=C_{\mu\nu\rho}\,l^\mu k^\nu l^\rho, \\
&\Psi_4:=C_{\mu\nu\rho}\,l^\mu m^\nu l^\rho,
\end{aligned}
\end{align}
where $C_{\mu\nu\rho}$ is the {\it Cotton tensor}~\cite{Cotton:1899}, and a null triad ${\{k^\mu, l^\mu, m^\mu\}}$ is properly normalized such that
\begin{align}\label{triad}
&k_\mu k^\mu=0=l_\mu l^\mu, \qquad k_\mu l^\mu=-1, \qquad m_\mu m^\mu=1, \qquad k_\mu m^\mu=0=l_\mu m^\mu.
\end{align}
Following the convention of~\cite{Garcia:2003bw}, we define the Cotton tensor as
\begin{align}
C_{\mu\nu\rho}:=2\,\big(\nabla_{[\mu}R_{\nu]\rho}-\tfrac14\nabla_{[\mu}R\,g_{\nu]\rho}\big),
\end{align}
which automatically satisfies the constraints ${C_{(\mu\nu)\rho}\equiv 0}$, ${C_{[\mu\nu\rho]}\equiv 0}$, and ${C_{\mu\nu}^{~~\mu}\equiv 0}$.
The Cotton scalars~$\Psi_{\rm A}$ are three-dimensional counterparts of the Newman-Penrose complex Weyl scalars of four-dimensional gravity~\cite{Newman:1961qr}. Moreover, for the algebraic classification, it is then convenient to employ the scalar polynomial invariants ~\cite{Podolsky:2023qiu,Papajcik:2023zen}
\begin{align}
\label{invariants}
\begin{aligned}
I:=&\ \Psi_0\Psi_4-2\Psi_1\Psi_3-3\Psi_2^2, \\
J:=&\ 2\Psi_0\Psi_2\Psi_4+2\Psi_1\Psi_2\Psi_3+2\Psi_2^3+\Psi_0\Psi_3^2-\Psi_4\Psi_1^2, \\
G:=&\ \Psi_1\Psi_4^2-3\Psi_2\Psi_3\Psi_4-\Psi_3^3,\\
H:=&\ 2\Psi_2\Psi_4+\Psi_3^2,\\
N:=&\ 3H^2+\Psi_4^2I.
\end{aligned}
\end{align}

As the metric (\ref{metric-y}) can be written as
\begin{align}
\label{metric-y-sq}
\D s^2 =&-\left(\sqrt{-\Lambda} \,\D t+\epsilon\,\D \theta\right)\!
\left[2\sqrt{-\Lambda}\,(y-D)\, \D t-S\,(\sqrt{-\Lambda} \,\D t+\epsilon\,\D \theta)\right] + \frac{\D y^2}{4(-\Lambda)\,(y-D)^2},
\end{align}
a natural triad $\{k_\mu, l_\mu, m_\mu\}$ in the coordinate system (\ref{metric-y}) satisfying \eqref{triad} is given by the one-forms
\begin{align}
\label{basis-CS-y-triad}
\begin{aligned}
k_\mu\,\D x^\mu=&-\frac{1}{\sqrt{2}}\left(\sqrt{-\Lambda} \,\D t+\epsilon\,\D \theta\right) ,\\
l_\mu\,\D x^\mu=&-\frac{1}{\sqrt{2}}\left(2\sqrt{-\Lambda}(y-D)\, \D t
-S\,(\sqrt{-\Lambda} \,\D t+\epsilon\,\D \theta)\right),\\
m_\mu\,\D x^\mu=&\ \frac{\D y}{2\sqrt{-\Lambda}\,(y-D)},
\end{aligned}
\end{align}
of which contravariant components are
\begin{align}
\label{basis-CS-y-triad-up}
\begin{aligned}
k^\mu\,\partial_\mu=&\ \frac{1}{\sqrt{2}\,(y-D)}\,\Big(\frac{1}{\sqrt{-\Lambda}}\,\partial_t-\epsilon\,\partial_\theta\Big) ,\\
l^\mu\,\partial_\mu=&\ \frac{1}{\sqrt{2}\,(y-D)}\,\Big(\frac{S}{\sqrt{-\Lambda}}\,\partial_t-\epsilon\,
\big[S-2(y-D)\big] \partial_\theta\Big) ,\\
m^\mu\,\partial_\mu=&\ 2\sqrt{-\Lambda}\,(y-D)\,\partial_y.
\end{aligned}
\end{align}
The null vector $k^\mu$ satisfies the geodesic conditions ${k^\nu\nabla_\nu k^\mu=0}$, while $l^\mu$ does not.
Also, as ${k_\mu=-(\nabla_\mu u)/\sqrt{2}}$ is satisfied, where $u$ is defined by Eq.~(\ref{defining-u}), $k^\mu$ is a normal vector of a null hypersurface given by $u=$constant.

The Cotton scalars \eqref{Cottonscalars} with respect to such a null triad are
\begin{align}\label{Cotton-CS-y}
\Psi_0 &= \Psi_1 = \Psi_2 = \Psi_3 =0,\qquad \Psi_4 = 16\pi G\, \sqrt{-\Lambda}\,{C_0^2},
\end{align}
which give the identically vanishing invariants,
\begin{align}
I=J=G=H=N=0.
\end{align}
Therefore, according to the flow diagram in Figure~1 in~\cite{Podolsky:2023qiu}, or equivalently in~\cite{Papajcik:2023zen}, the {\it CCS spacetime is of the Cotton type N everywhere} (unless we consider the vacuum case ${C_0=0}$ which is conformally flat, that is of type O).
This is in {striking contrast} to the large class of charged rotating BTZ black holes which are of Cotton type~I away from the horizon and type~III on the horizon. (See Corollary 1 in \cite{Maeda:2023oei}.)

As generally explained in detail in \cite{Papajcik:2023zen}, by a suitable choice of the triad the Cotton scalars for any type~N spacetime can be put into the canonical form in which {\it only the scalar $\Psi_4$ is nonzero}. It is the case of Eq.~\eqref{Cotton-CS-y}, and this explicitly demonstrates that the null vector ${k}^\mu\,\partial_\mu$ given by \eqref{basis-CS-y-triad-up} is the {\it quadruple Cotton-aligned null direction} (CAND) of the type N CCS spacetime. Moreover, \eqref{basis-CS-y-triad-up} is the {\it principal null triad}. (See~\cite{Podolsky:2023qiu,Papajcik:2023zen} for more details on the definition and multiplicity of CAND.)

Analogous (Newman-Penrose) scalars for the Maxwell field are computed to give
\begin{align}
\label{NP-Max2}
\begin{aligned}
\phi_0:=&\,F_{\mu\nu}\, {k}^\mu \,{m}^\nu = \,0,\\
\phi_1:=&\,F_{\mu\nu}\, {k}^\mu \,{l}^\nu \ \,= \,0,\\
\phi_2:=&\,F_{\mu\nu}\, {m}^\mu \,{l}^\nu \,= \epsilon \sqrt{2}\,C_0.
\end{aligned}
\end{align}
It proves that the {\it electromagnetic field is aligned} with the gravitational field (because ${\phi_0=0}$), and it is {\it null}, that is {\it radiative} (because ${\phi_0=0=\phi_1}$ but ${\phi_2\ne0}$). Expressed geometrically, the double-degenerate null direction of the type~N Maxwell field coincides with the quadruple-degenerate Cotton aligned null direction of the CCS type~N gravitational field. In fact, it was observed in \cite{Garcia-Diaz:2017cpv} that {\it both} the Cotton and Maxwell tensors possess the same triple zero eigenvalues, so that their algebraic types are N. (See page~188.)
However, these key properties of the {\it algebraically most special} (``null'', that is ``radiative'') type have not been taken into account for the physical (re)interpretations of the CCS solution.

The geometrically privileged null vector field ${{k}^\mu}$ of the principal triad \eqref{basis-CS-y-triad-up} satisfies the geodesic conditions (${{k}^\nu\nabla_\nu {k}^\mu=0}$).
Then, its {\it optical scalars are all zero}, in particular the expansion ${{\rho_{k}:=(\nabla_\mu{k}_\nu)\,{m}^\mu {m}^\nu }}$,
\begin{align}
\rho_{k}=0.
\end{align}
It means that the whole CCS family of solutions, including the Kamata-Koikawa solution, belongs to the {\it Kundt class} of spacetimes~\cite{Podolsky:2021gsa}.

\subsection{The Kundt canonical form of the solution}
\label{Sec:Kundt-canonical}
We have thus shown that the CCS spacetime belongs to the Kundt class in three dimensions. Now we write the metric of this solution in the canonical Kundt form.

Using a null coordinate $u$ defined by (\ref{defining-u}) instead of $t$, and renaming the spatial coordinate~$r$ as ${r \mapsto x}$, we write the metric (\ref{metric-rttheta}) and the gauge field (\ref{CSKK-A}) in the new coordinates $(u, x,\theta)$ as
\begin{align}
\label{metric-rutheta-xPH}
\begin{aligned}
&\D s^2 = \frac{\D x^2}{P^2} +2 \epsilon\,(x^2-D)\, \D u\,\D \theta + H\,\D u^2 ,\\[2mm]
&A_\mu\, \D x^\mu = \frac{\epsilon\,C_0}{2\sqrt{-\Lambda}} \ln|x^2-D|\,\D u,
\end{aligned}
\end{align}
where
\begin{align}
P(x)& := \sqrt{-\Lambda}\,\, \frac{x^2-D}{x}, \\
H(x)& := 2D-x^2+\frac{\kappa{C_0^2}}{2(-\Lambda)}\ln|x^2-D|.
\end{align}
The Faraday tensor is given by
\begin{align}
\mathbf{F} = F_{\mu\nu}\,\D x^\mu\wedge\D x^\nu
= \frac{\epsilon\,C_0}{\sqrt{-\Lambda}}\, \frac{x}{x^2-D}\, \D x \wedge \D u.\label{F-uxtheta}
\end{align}
This is a three-dimensional analogue of the four-dimensional metric representing a family of {\it all} type-N Kundt spacetimes which are the solutions in the Einstein-Maxwell-$\Lambda$ system, or in the system with a null dust fluid instead of the Maxwell field.
In four dimensions, such the most general type~N Kundt solution was first presented in~\cite{Ozsvath:1985qn}, and later investigated in detail~\cite{Bicak:1999ha}. (See Chap.~18 of the textbook~\cite{Griffiths:2009dfa} for a review.)

Finally, by introducing a new coordinate $r_{\rm\!_K}$, instead of $\theta$, defined by
\begin{align}\label{r-Kundt}
r_{\rm\!_K} : = -\epsilon\,(x^2-D)\,\theta,
\end{align}
the CCS metric (\ref{metric-rutheta-xPH}) is written in the {\it canonical Kundt form} in the coordinates $(u,x,r_{\rm\!_K})$ as
\begin{align}
\D s^2 = \frac{\D x^2}{P^2} + \frac{4\,x\,r_{\rm\!_K}}{x^2-D}\, \D u\,\D x -2\,\D u\,\D r_{\rm\!_K} + H\,\D u^2, \label{metric-Kundt}
\end{align}
while $A_\mu$ and ${\bf F}$ are unchanged.
It actually belongs to a special family of {\it degenerate Kundt metrics}. (See, e.g., Sec.~7.1 of the topical review~\cite{OrtaggioPravdaPravdova:2013} for the definition and more details.)

The metric (\ref{metric-Kundt}) can be directly compared with the {complete family} of three-dimensional Kundt solutions with a Maxwell field (necessarily aligned) and $\Lambda$, found recently in \cite{Podolsky:2021gsa}.
Comparing the metric functions in Eq.~\eqref{metric-Kundt} with the general expressions ${g_{ux}=e+f\,r_{\!_K}}$ and ${g_{uu}=a+b\,r_{\!_K}+c\,r_{\!_K}^2}$ given in Eq.~(90) in~\cite{Podolsky:2021gsa}, we identify
\begin{align}
\label{metric-2022-PRD-paper}
\begin{aligned}
& e=0,\qquad f = \frac{2x}{x^2-D}\ne0,\qquad a=H,\\
& b=0,\qquad c = - \tfrac{1}{2}\kappa_0Q^2 = 0\quad\Rightarrow\ Q=0 .
\end{aligned}
\end{align}
The corresponding Maxwell field, given by Eq.~(95) in~\cite{Podolsky:2021gsa}, is thus ${\mathbf{F} = \xi(x)\, \D x \wedge \D u}$, that is easily identified with Eq.~\eqref{metric-rutheta-xPH}. It is an aligned null Maxwell field that belongs to the subcase~(i) with ${\phi_0=0=\phi_1}$. (See Eq.~(99) in~\cite{Podolsky:2021gsa}.)
Moreover, by performing the transformations \eqref{defining-u} and \eqref{r-Kundt}, the principal null triad \eqref{basis-CS-y-triad-up} becomes
\begin{align}
\label{prinipal-triad5}
{k}^\mu\,\partial_\mu =\frac{1}{\sqrt{2}}\,\partial_{r_{\rm\!_K}}, \qquad
{l}^\mu\,\partial_\mu = \frac{1}{\sqrt{2}}\,H\,\partial_{r_{\rm\!_K}} +\sqrt{2} \partial_u , \qquad
{m}^\mu\,\partial_\mu = P\,(\partial_x +g_{ux}\, \partial_{r_{\rm\!_K}}).
\end{align}
After an additional simple boost $B\,{k}^\mu\to {k}^\mu$ and $B^{-1}\,{l}^\mu\to {l}^\mu$ with ${B=\sqrt{2}}$, the triad (\ref{prinipal-triad5}) fully agrees with Eq.~(6) in~\cite{Podolsky:2021gsa}.

It can also be observed that by introducing a modified spatial coordinate
\begin{align}
\label{def-z}
z : = \frac{1}{2\sqrt{-\Lambda}}\,\ln|x^2-D|,
\end{align}
the metric \eqref{metric-Kundt} of the CCS solution is put into the form
\begin{align}
\label{metric-with-z}
\D s^2 = \D z^2 + 4\,\sqrt{-\Lambda}\,\,r_{\rm\!_K}\,\D u\,\D x -2\,\D u\,\D r_{\rm\!_K}
+
\left(\,D - e^{2\,\sqrt{-\Lambda}\,z} + \frac{\kappa{C_0^2}}{\sqrt{-\Lambda}}\,z \,\right)\D u^2,
\end{align}
in which the p.p. curvature singularity (originally at ${r^2=D}$) is located at ${z\to-\infty}$. In this coordinate system $(u,z,r_{\rm\!_K})$, the Maxwell field (\ref{F-uxtheta}) is {\it uniform}, namely
\begin{align}
\label{elmag-with-z}
\mathbf{F} = \epsilon\,C_0\, \D z \wedge \D u = \frac{\phi_2}{\sqrt{2}} \, \D z \wedge \D u.
\end{align}

It should finally be noted that the physical interpretation of the family of closely related four-dimensional Kundt spacetime of algebraic type N with any $\Lambda$ was investigated in \cite{GriffithsDochertyPodolsky:2004, Podolsky:2004qu}. It elucidated the character of wave surfaces in these spacetimes, and also the related p.p. singularity. It is the {\it caustics} formed as an envelope of the wave surfaces (see Chap.~18 of the textbook~\cite{Griffiths:2009dfa} for a review.)

\subsection{Compatible matter fields}

Lastly, we show that the CCS solution (\ref{metric-y}) can be a solution not only with the original Maxwell field but also with a {\it null dust fluid} or a {\it massless scalar field}.
For this purpose, we study the Hawking-Ellis type of $T_{\mu\nu}$ for the solution.
With the orthonormal basis one-forms given by
\begin{align}
\label{basis-T}
&E_\mu^{(0)}=\frac{1}{\sqrt{2}}(l_\mu+k_\mu),\qquad E_\mu^{(1)}=\frac{1}{\sqrt{2}}(l_\mu-k_\mu),\qquad E_\mu^{(2)}=m_\mu
\end{align}
constructed from a triad $\{k_\mu, l_\mu, m_\mu\}$ in Eq.~(\ref{basis-CS-y-triad}), non-zero components of $T^{(a)(b)}:=T^{\mu\nu}E_\mu^{(a)}E_\nu^{(b)}$ are computed to give
\begin{align}
T^{(0)(0)} = \,T^{(1)(1)} = \, T^{(0)(1)}\,\big(=T^{(1)(0)}\,\big) =
C_0^2, \label{Tab-sol}
\end{align}
which shows
\begin{align}
&\big(\,T^{(0)(0)}+T^{(1)(1)}\big)^2-4(\,T^{(0)(1)})^2=0.
\end{align}
Then, by Lemma~1 in~\cite{Maeda:2022vld}, $T_{\mu\nu}$ for the Maxwell field in the CCS solution is of the Hawking-Ellis type II everywhere.

Additionally, the expression (\ref{Tab-sol}) of $T^{(a)(b)}$ shows that the CCS metric also solves the Einstein-$\Lambda$ equations with a null dust fluid, of which energy-momentum tensor is given by
\begin{align}
\label{null-sol}
T_{\mu\nu}=\Omega \,k_\mu k_\nu.
\end{align}
Here $k_\mu$ is the null vector in Eq.~(\ref{basis-CS-y-triad}) and the energy density of the null dust $\Omega$ is constant given by
\begin{align}
\label{null-sol2}
\Omega = 2\,C_0^2.
\end{align}

In three dimensions, the field equations (\ref{EFE}) are equivalent to the ones with a massless scalar field $\phi$ instead of the Maxwell field, using the duality.
The dual Maxwell one-form is defined by
\begin{align}
{}^*F_{\mu}:=\frac12\varepsilon_{\mu\nu\rho}\,F^{\nu\rho}~~~\left(\ \Leftrightarrow {}^*F_{\mu}\,\varepsilon^{\mu\alpha\beta}=-F^{\alpha\beta}\right),\label{dualF}
\end{align}
where the totally anti-symmetric volume three-form $\varepsilon_{\mu\nu\rho}$ is defined by
\begin{align}
\varepsilon_{\mu\nu\rho}:=\sqrt{-g}\,\epsilon_{\mu\nu\rho}~~~\left(\Leftrightarrow \varepsilon^{\mu\nu\rho}=-\epsilon^{\mu\nu\rho}/\sqrt{-g}\right)
\end{align}
with the Levi-Civita symbol $\epsilon_{\mu\nu\rho}$ satisfying $\epsilon_{012}=1$ and $\epsilon^{012}=1$.
Identifying
\begin{align}
\nabla_\mu \phi\equiv {}^*F_{\mu},
\end{align}
and using $\varepsilon^{\mu\alpha\beta}\varepsilon_{\mu\nu\rho}=-(\delta^\alpha_{~\nu}\delta^\beta_{~\rho}-\delta^\alpha_{~\rho}\delta^\beta_{~\nu})$, we obtain the equation of motion and the energy-momentum tensors of $\phi$ as
\begin{align}
\nabla^2\phi=&\ 0,\\
T_{\mu\nu}=&\ F_{\mu\rho}F_\nu^{~\rho}-\frac 14 g_{\mu\nu}F_{\rho\sigma}F^{\rho\sigma} \nonumber\\
=&\ (\nabla_{\mu}\phi)(\nabla_{\nu}\phi)-\frac 12g_{\mu\nu}(\nabla\phi)^2, \label{Tab-dual}
\end{align}
where ${\nabla^2\phi:=\nabla_\mu\nabla^\mu\phi}$ and ${(\nabla\phi)^2:=(\nabla_{\alpha}\phi)(\nabla^{\alpha}\phi)}$.

For the CCS solution, the dual massless scalar field is given in the coordinate system (\ref{metric}) as
\begin{align}
\phi=C_0\,\big(\sqrt{-\Lambda}\,t + \epsilon\,\theta\,\big)+\phi_0 = C_0\,u+\phi_0,\label{phi-CS}
\end{align}
where $\phi_0$ is a constant, and $u$ is the null coordinate defined by Eq.~(\ref{defining-u}).
Different from the case with a Maxwell field or a null dust fluid, the coordinate $\theta$ now cannot be periodic, as a periodic boundary condition ${\phi(t,\theta)=\phi(t,\theta+2\pi)}$ is not satisfied unless ${C_0=0}$.

\section{Concluding remarks}
\label{sec:summary}
In our paper, we have studied in detail the CCS solution (\ref{metric-y}) with the metric function~(\ref{def-S}) in the coordinates $(t,y,\theta)$. It is the known most general stationary and axisymmetric solution in the three-dimensional Einstein-Maxwell-$\Lambda$ system under the so-called {\it self-dual} or {\it anti-self-dual} condition on the Maxwell field~\cite{Cataldo:1999fh}. The solution is locally characterized by a single parameter $C_0$, and it reduces to the extremal rotating BTZ vacuum solution for ${C_0=0}$. Nevertheless, we have kept an auxiliary parameter $D$ which allows the limit to the Kamata-Koikawa solution.

Our results for the {\it charged} case ${C_0\ne 0}$ are summarized as follows.
\begin{enumerate}

\item For any value of $D$, the domain of the radial coordinate is ${y\in(D,\infty)}$. The spacetime is asymptotically locally AdS near the spacelike and null infinities given by ${y\to \infty}$. At ${y=D}$, there is a p.p. curvature singularity which corresponds to an infinite affine parameter along specific spacelike geodesics.

\item If $\theta$ is a periodic coordinate such that $(t,y,\theta)=(t,y,\theta+2\pi)$, there are closed timelike curves near the singularity ${y=D}$.

\item The spacetime is of the Cotton type~N everywhere. We have identified a uniquely chosen principal null triad and the quadruple Cotton-aligned null direction (CAND).

\item The Maxwell field is null, namely radiative. The double-degenerate null direction of the type~N Maxwell field coincides with the quadruple-degenerate CAND of the type~N gravitational field.

\item We have written the CCS solution in the canonical Kundt form and identified the privileged null coordinate $u$ given by Eq.~(\ref{defining-u}) such that ${u=u_0=\,}$constant labels privileged null hypersurface in the Kundt family.

\item The energy-momentum tensor of the solution is of the Hawking-Ellis type II everywhere. The CCS metric also solves the three-dimensional Einstein-$\Lambda$ equations with a null dust fluid given by Eqs.~(\ref{null-sol}) and (\ref{null-sol2}) or a massless scalar field given by Eq.~(\ref{phi-CS}).

\end{enumerate}
To summarize, the CCS solution is algebraically, geometrically, and physically {\it different} from the charged rotating BTZ solution~\cite{Clement:1995zt} and does {\it not} describe a black hole, unless the Maxwell field is trivial.

In the three-dimensional Einstein-Maxwell-$\Lambda$ system, many stationary and axisymmetric symmetric solutions have been classified and described by Garc\'\i{}a-D\'\i{}az~\cite{Garcia-Diaz:2013aaa,Garcia:2009vsk}.
However, to the best of the authors' knowledge, it is still an open problem whether there is a different class of charged rotating asymptotically AdS black-hole solutions other than the charged rotating BTZ solution. This problem is left for future investigation.

\section*{Acknowledgments}
H.M. is very grateful to Max-Planck-Institut f\"ur Gravitationsphysik (Albert-Einstein-Institut) and the Institute of Theoretical Physics in Prague (Charles University) for their hospitality.
This work has been supported by the Czech Science Foundation Grant No.~GA\v{C}R 23-05914S.

\appendix

\section{Cl\'ement's solution in the 1993 paper}
\label{app:Clement}

In this appendix, we show that Cl\'ement's solution given by Eq.~(29) in Ref.~\cite{Clement:1993kc} with the Minkowski signature $(+,-,-)$ is locally identical to the Cataldo-Salgado solution.\footnote{Note that (23) in Ref.~\cite{Clement:1993kc} is the charged rotating BTZ solution after the double Wick rotation.}
The original solution can be written in the coordinates $(t,\rho,\theta)$ with the Minkowski signature $(-,+,+)$ as
\begin{align}
\label{Clement-app}
\begin{aligned}
&\D s^2=\pm\frac{4}{\ell}\,\rho\,\D t\,\D\theta
  +\big(-b\,\rho+2\kappa q^2\ln|\rho/\rho_0|\big)\,\D\theta^2+\frac{\ell^2}{4\rho^2}\,\D\rho^2,\\
&A_\mu\D x^\mu=q\ln|\rho/\rho_0|\D\theta,
\end{aligned}
\end{align}
where $\rho_0$, $b$, and $q$ are constants, and $\ell$ is the AdS radius defined by $\ell=1/\sqrt{-\Lambda}$.

The Cataldo-Salgado solution in the single-null coordinates (\ref{metric-yu}) is
\begin{align}
\label{metric-yu-app}
\begin{aligned}
&\D s^2 = -\Bigl(y-2D-\frac{\kappa{C_0^2}}{2(-\Lambda)}\ln|y-D|\Bigl)\,\D u^2+2 \epsilon\,(y-D)\, \D u\,\D \theta + \frac{\D y^2}{4(-\Lambda)(y-D)^2},\\
&A_\mu\,\D x^\mu =\frac{\epsilon\,C_0}{2\sqrt{-\Lambda}} \ln|y-D|\,\D u.
\end{aligned}
\end{align}
By coordinate transformations
\begin{align}
y-D=\alpha\, \rho,\qquad u=\beta\,{\bar\theta},\qquad \theta=(\alpha \beta)^{-1}\,\frac{2}{\ell}\, t
\end{align}
with constant $\alpha$ and $\beta$ satisfying
\begin{align}
\frac{\kappa{C_0^2}}{2(-\Lambda)}\ln|\alpha|=-D-\frac{\kappa{C_0^2}}{2(-\Lambda)}\ln|\rho_0|,\qquad
\beta^2=\frac{b}{\alpha}
\end{align}
and using the freedom of a gauge constant of $A_\mu$, the solution (\ref{metric-yu-app}) is transformed into Cl\'ement's solution (\ref{Clement-app}) with $q=\epsilon\beta\,C_0/(2\sqrt{-\Lambda})$.

\section{Parameters of the {Cataldo-Salgado} solution}
\label{app:CSKK}

Cataldo and Salgado presented their solution as
\begin{align}
\label{Cataldo-Salgado}
\begin{aligned}
&\D s^2=-\frac{r^2}{R^2}\,f\,\D t^2+\frac{\D r^2}{f}+R^2\Big(\D\theta+\frac{h}{R^2}\,\D t\Big)^2,\\
&A_\mu\D x^\mu=\frac{C_0}{2\sqrt{-\Lambda}}\ln\biggl|\frac{r^2-D}{C}\biggl|\Big[\Big(\epsilon\sqrt{-\Lambda}+\frac{E}{C}\Big)\D t+\D\theta\Big]
\end{aligned}
\end{align}
with the metric functions
\begin{align}
f(r)&=(-\Lambda)\,\frac{(r^2-D)^2}{r^2},\\
R(r)&=\sqrt{r^2+\frac{\kappa C_0^2}{2(-\Lambda)}\ln\biggl|\frac{r^2-D}{C}\biggl|},\\
h(r)&=\frac{E}{C}\,r^2 +\epsilon D\sqrt{-\Lambda} +\frac{\kappa {C_0^2}}{2\sqrt{-\Lambda}}
\Big(\epsilon+\frac{E}{\sqrt{-\Lambda}\,C}\Big)\ln\biggl|\frac{r^2-D}{C}\biggl|. \label{h-appendix}
\end{align}
See Eqs. (37)--(39) in Ref.~\cite{Cataldo:1999fh}, in which ${C\,(\ne 0)}$, $D$, $C_0$ and $E$ are four arbitrary constants of integration, and ${\epsilon=\pm 1}$.

However, without loss of generality we can {\it locally} set ${C=1}$ and ${E=0}$ by the following coordinate transformations
\begin{align}
t = \frac{\bar t}{\sqrt{C}}, \qquad
r = \sqrt{C}\,{\bar r},\qquad
\theta = \frac{{\bar \theta}-{\bar E}\,{\bar t}}{\sqrt{C}}
\end{align}
with the reparametrizations ${E= C{\bar E}}$, ${D= C{\bar D}}$, and ${C_0= \sqrt{C}{\bar C}_0}$.
Nevertheless, the constants $C$ and $E$ may possibly have global meanings.
For example, if one assumes that $\theta$ is periodic as $\theta\in[0,2\pi)$ in the coordinates (\ref{Cataldo-Salgado}) with ${C\ne 1}$ and ${E\ne 0}$, then ${{\bar\theta}=0}$ is not identified with ${{\bar\theta}=2\pi}$ at different times.

With ${C=1}$ and ${E=0}$, using the useful identities (\ref{CSKK-functions-identities}) in the region of ${r^2>D}$, the Cataldo-Salgado solution can be rewritten in the simple form (\ref{metric-rttheta}),
\begin{align}
\label{metric-rttheta-app}
\begin{aligned}
&\D s^2 = (-\Lambda)\big[R^2-2(r^2-D) \big]\,\D t^2 + 2\epsilon\sqrt{-\Lambda}\,\big[R^2-(r^2-D) \big]\, \D t\,\D \theta+R^2\D \theta^2 +\frac{\D r^2}{f},\\
&A_\mu\,\D x^\mu =\frac{1}{2}\,C_0\,\ln|r^2-D|\,\Big(\epsilon\,\D t + \frac{\D\theta}{\sqrt{-\Lambda}}\Big),
\end{aligned}
\end{align}
where $R^2$ is given by Eq.~(\ref{CSKK-functions}). Changing the sign of $\epsilon$ is equivalent to a transformation of time reversal ${t\to -t}$. Then, without loss of generality we can also {\it locally} set ${D=0}$ in the charged case $C_0\ne 0$ by the following coordinate transformations
\begin{align}
&r^2-D=e^{-2(-\Lambda)D/(\kappa{C_0^2})} \,{\bar r}^2, \qquad
 t=e^{(-\Lambda)D/(\kappa{C_0^2})} \,{\bar t},\qquad
 \theta=e^{(-\Lambda)D/(\kappa{C_0^2})}\,{\bar\theta}
\end{align}
with a reparametrization ${C}_0=e^{-(-\Lambda)D/(\kappa{C_0^2})}{\bar C}_0$, which generate a gauge constant for $A_\mu$.
However, setting $D=0$ forces us to miss the limit from the Cataldo-Salgado solution (\ref{metric-rttheta-app}) to the Kamata-Koikawa solution~\cite{Kamata:1995zu} for ${\kappa {C_0^2}=-2\Lambda D}$.

\end{document}